\documentclass[journal]{IEEEtran}

\usepackage{amsmath}
\usepackage{lipsum}
\usepackage{graphicx}
\usepackage[justification=centering]{caption}
\usepackage{epstopdf}
\usepackage{tabularx}
\usepackage{algorithm}
\usepackage{algorithmic}

\usepackage{multirow}
\usepackage{xcolor}

\usepackage{makecell}
\ifCLASSINFOpdf
\else
\fi

\usepackage{color,soul}

\begin{document}
	%
	
	\title{IoT-Enabled Social Relationships Meet Artificial Social Intelligence}
	
	%
	%
	
	\author{Michael~Shell,~\IEEEmembership{Member,~IEEE,}
		John~Doe,~\IEEEmembership{Fellow,~OSA,}
		and~Jane~Doe,~\IEEEmembership{Life~Fellow,~IEEE}
		\thanks{M. Shell was with the Department
			of Electrical and Computer Engineering, Georgia Institute of Technology, Atlanta,
			GA, 30332 USA e-mail: (see http://www.michaelshell.org/contact.html).}
		\thanks{J. Doe and J. Doe are with Anonymous University.}
		\thanks{Manuscript received April 19, 2005; revised August 26, 2015.}}

\author{Sahraoui Dhelim, Huansheng Ning, Fadi Farha, Liming Chen, Luigi Atzori and Mahmoud Daneshmand
	\thanks{Sahraoui Dhelim, Huansheng Ning and Fadi Farha are with the School of Computer and Communication Engineering, University of Science and Technology Beijing, 100083, Beijing, China.}
	\thanks{Sahraoui Dhelim and Huansheng Ning are also with Beijing Engineering Research Center for Cyberspace Data Analysis and Applications, Beijing, China}
	\thanks{Liming Chen is with the School of Computing, Ulster University, Newtownabbey BT37 0QB, United Kingdom}%
	\thanks{Luigi Atzori is with the Department of Electrical and Electronic Engineering, University of Cagliari, piazza d’Armi, 09123 Cagliari, Italy}
	\thanks{Mahmoud Daneshmand is with the Department of Business Intelligence and Analytics and the Department of Computer Science, Stevens Institute of Technology, Hoboken, USA.}
	\thanks{Corresponding author: Huansheng Ning (ninghuansheng@ustb.edu.cn).}
	\thanks{© 2021 IEEE.  Personal use of this material is permitted.  Permission from IEEE must be obtained for all other uses, in any current or future media, including reprinting/republishing this material for advertising or promotional purposes, creating new collective works, for resale or redistribution to servers or lists, or reuse of any copyrighted component of this work in other works.}
}

	%
	%

	\markboth{IEEE Internet of Things Journal , 2021}%
	{Shell \MakeLowercase{\textit{et al.}}: Bare Demo of IEEEtran.cls for IEEE Journals}
	%



	\maketitle
	
	\begin{abstract}
		With the recent advances of the Internet of Things, and the increasing accessibility to ubiquitous computing resources and mobile devices, the prevalence of rich media contents, and the ensuing social, economic, and cultural changes, computing technology and applications have evolved quickly over the past decade. They now go beyond personal computing, facilitating collaboration and social interactions in general, causing a quick proliferation of social relationships among IoT entities. The increasing number of these relationships and their heterogeneous social features have led to computing and communication bottlenecks that prevent the IoT network from taking advantage of these relationships to improve the offered services and customize the delivered content, known as social relationships explosion. On the other hand, the quick advances in artificial intelligence applications in social computing have led to the emerging of a promising research field known as Artificial Social Intelligence (ASI) that has the potential to tackle the social relationships explosion problem. This paper discusses the role of IoT in social relationships management, the problem of social relationships explosion in IoT, and reviews the proposed solutions using ASI, including social-oriented machine-learning and deep-learning techniques.
	\end{abstract}
	
	
	%
	\IEEEpeerreviewmaketitle

\section{Introduction}

\IEEEPARstart{W}{ith} the fast development of the Internet of Things (IoT), the massive proliferation of connected devices is expected to reach 41 billion devices connected to the IoT network by 2025 \cite{IDC2019}. The IoT network is facing various scalability challenges. The most prominent problems are known as the scalability explosions. The IoT is suffering from data explosion at the sensing layer \cite{Verma2017}, connections explosion at the network layer \cite{Ning2020}, and application/services explosion at the applications layer \cite{ning2020iot}. The nature of IoT devices has changed over the years. In the next generation of IoT, the objects are integrated with our social dimension, making them smart and social objects \cite{du2018social}. The integration of physical devices with the users' social dimension has enabled them to understand the social context of the users and perform a whole new type of social computing tasks \cite{M.S.2019}. On the other hand, the popularity of online social networks has led to the emergence of a new type of social networking application that could operate at the Unit IoT level \cite{ning2013unit}. As a result of that, we have witnessed a quick proliferation of social relationships among IoT entities, such as user-user relationship, user-device relationship, and device-device relationship. These relationships are empowering IoT applications with key functionalities, such as social trust analysis, users' social attributes profiling, management of social communities, and social recommendation services. However, the increasing number of these relationships and their heterogeneous social features have led to a computing and communication bottleneck that prevents the IoT network from taking advantage of these relationships to improve the offered services and customize the delivered content. This is known as social relationships explosion. 

\begin{figure}[htb!]
	\centering
	\includegraphics[width=0.4\columnwidth]{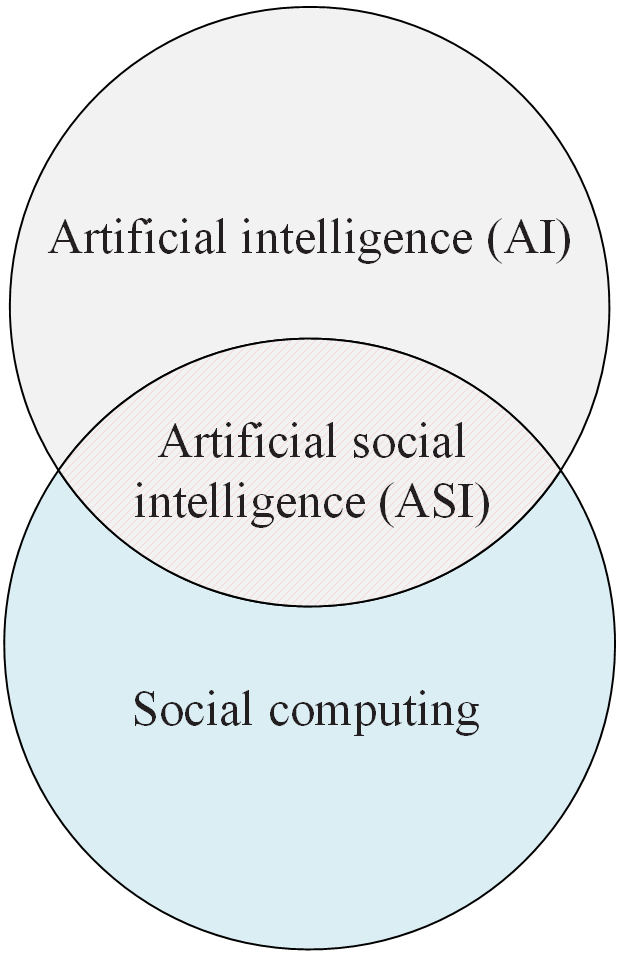}
	\caption{The scope of artificial social intelligence}
	\label{asi}
\end{figure}

In recent years, we have witnessed a new computing paradigm known as social computing, which focuses on integrating the social dimension in the computing system. Social computing aims to enable smart systems to socialize with the user and understand its social context. Moreover, the quick advances in Artificial Intelligence (AI) application in social computing have led to an emerging promising study field known as Artificial Social Intelligence (ASI). It has emerged as a result of the joint study area shared between AI and social computing, see Figure \ref{asi}. ASI has the potential to tackle the social relationships explosion, as it approaches this problem from a social computing perspective, unlike conventional AI.

\begin{figure}[htb!]
	\centering
	\includegraphics[width=0.7\columnwidth]{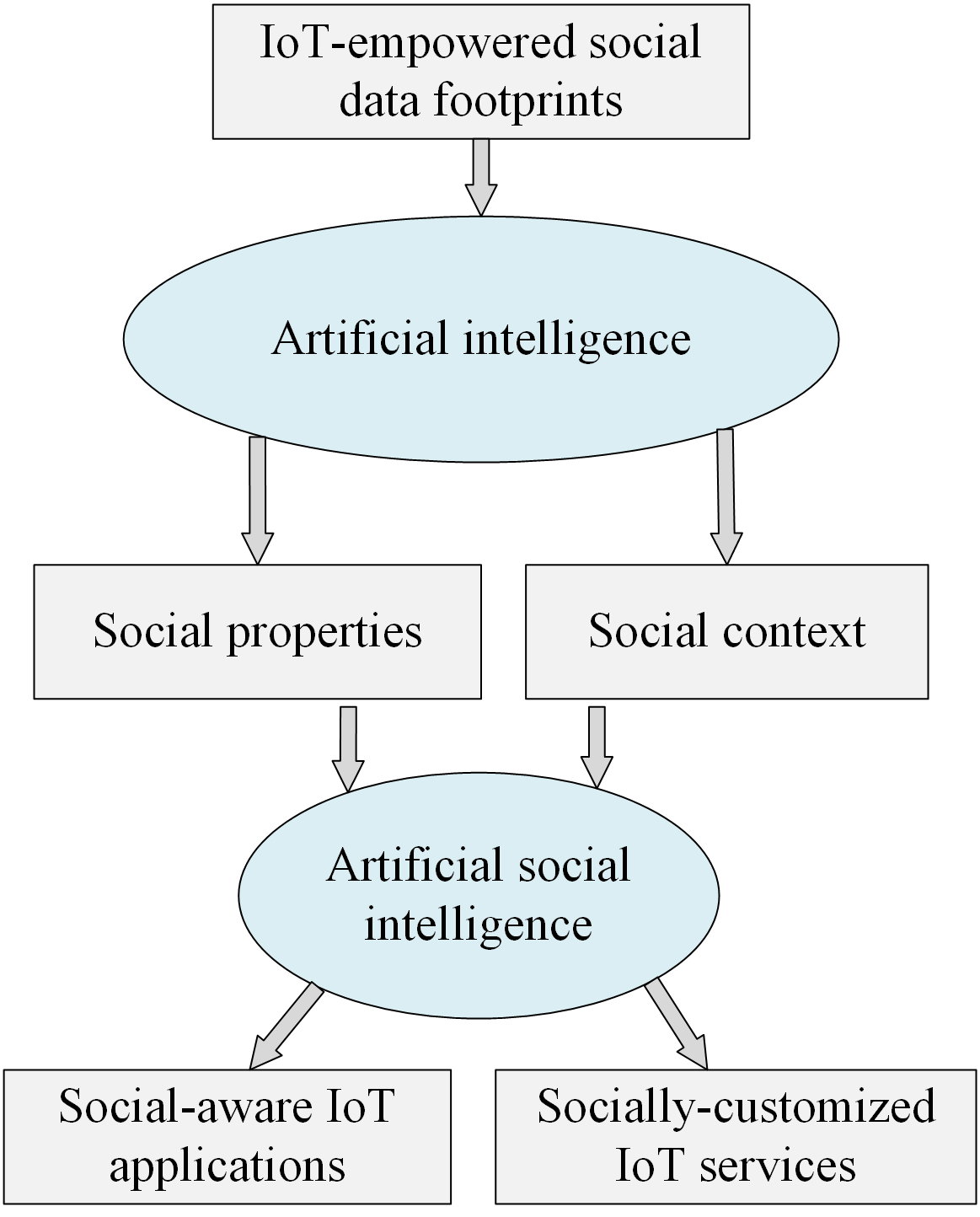}
	\caption{Social footprint processing flow}
	\label{footprint}
\end{figure}

The enormous volume of data  generated by IoT devices can be a rich source of the user's social properties. By analyzing the users' generated social data footprints recorded from a different situation of physical contact and communication among users (e.g., location, emails log, call records and text messages), the emotional states and behavioral patterns, the smart devices can be aware of the physical as well as the social context of the users. Accordingly, they will offer personalized services and customized content according to their social context. The conventional AI data analysis technologies should work simultaneously with the ASI to maximize the system knowledge about the social dimension of the users. As shown in Figure \ref{footprint}, applying conventional AI on the user's social data footprint empowered by IoT systems can help to extract social features and social context data. Moreover, ASI is applied to understand the social context's semantic and customize the services and applications accordingly.

This paper discusses the role of IoT in social relationships detection and management and the problem of social relationships explosion in IoT, and reviews the proposed solutions using ASI, including social-oriented machine-learning and deep-learning techniques.

The remainder of this paper is organized as follows: Section \ref{sec.r1} highlights the existing research and the gaps in existing research on the topic. Section \ref{sec.2} argues the necessity of using ASI and the main differences between conventional AI and ASI. Section \ref{sec.3} examines the different types of relationships that connects the IoT entities. Section \ref{sec.4} discusses some of the challenges that face the IoT networks due to social relationships explosion. Section \ref{sec.5} reviews some of the works that proposed social-aware approaches for IoT. Section \ref{sec.r2} presents a use case scenario of an ASI application. Section \ref{sec.6} presents some of the ASI-enabled IoT applications. Finally, Section \ref{sec.7} concludes the paper.

\section{Related work}
\label{sec.r1}

Atzori \textit{et al.} \cite{Atzori2012} proposed the integration of social network with IoT to form the SIoT network, where the SIoT architecture can guarantee the network navigability and the discovery of objects and services is performed effectively and the scalability is ensured just like in the human social networks. The advantage of the proposed SIoT paradigm is that the social networks can be re-used to solve IoT related issues related to the scalability of interconnected objects. To further explain the benefits of SIoT compared to the conventional IoT. The same research group \cite{Atzori2014} presented the advantages of interconnecting 'social objects' rather than 'smart objects', which is considered a generational leap from objects with a certain degree of smartness to objects with an actual social consciousness. Similarly, in another work \cite{atzori2011siot} they discussed the SIoT architectural model, and discussed the various social relationships that can connect the objects of SIoT network. Wang \textit{et al.} \cite{Wang2007} discussed the potentials of social computing paradigm, and show that social computing technologies move beyond social information processing towards emphasizing social intelligence. Specifically, they proved that the move from social informatics to social intelligence is achieved by modeling and analyzing social features and behaviors, and by capturing human social dynamics, and by creating artificial social agents and generating and managing actionable social knowledge within the IoT network. Recently, Khelloufi \textit{et al.} \cite{Khelloufi2020} proposed a service recommendation system that leverages the social relationships between IoT devices' owners, where the recommendation is based on the different relationships between the service requester and service provider, furthermore, they proposed a boundary based community detection algorithm that we used to form socially-connected device communities.

While all the above-mentioned works have discussed the importance of incorporating the social relationships and objects' social  properties in IoT computing and communication schemes. None of these works have addressed the role of AI in IoT-enabled social computing. The current paper advocates for considering the social properties and attributions of IoT entities through the application of artificial social intelligence.

\section{Relationships in IoT}
\label{sec.2}

\begin{figure*}[htb!]
	\centering
	\includegraphics[width=0.8\textwidth,height=7cm]{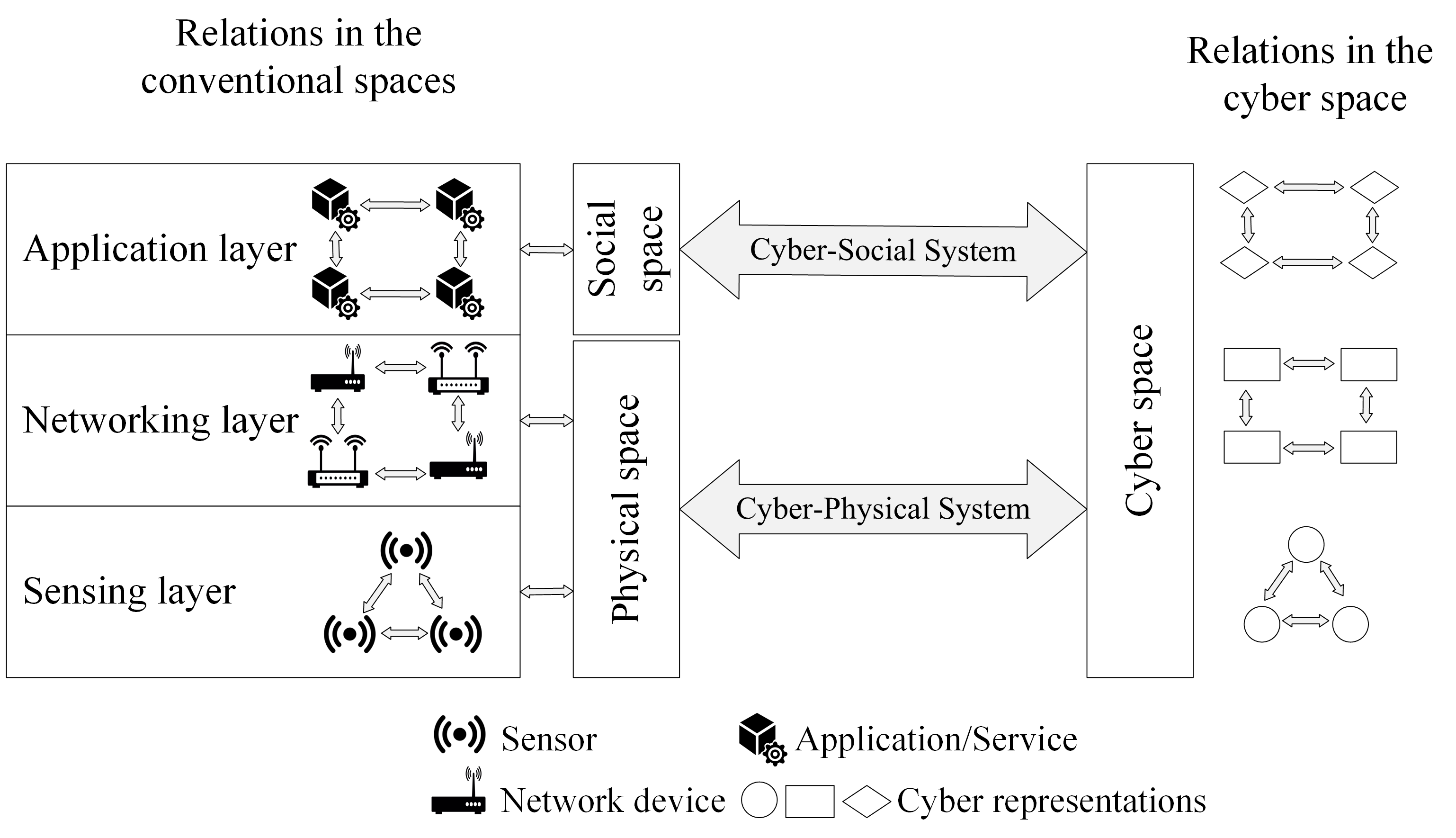}
	\caption{Relationships cyber mapping}
	\label{cyber_mapping}
\end{figure*}

As stated early, the future IoT network will incorporate billions of connected sensors, actuators and devices, along with networking equipment, and millions of applications and connected users. All these entities are interconnected by various types of relationships that describe the common functionality shared by these entities. Manipulating the properties of these relationships to adapt to the continuous changes in the surrounding environment enable the system to optimize the performance of these entities without the need for a change in the physical hardware. There are different types of relationships that interconnect IoT entities:
 
Hierarchical relationship \cite{zhu2020huna}: defines the hierarchical relationship that connects two IoT entities. Two entities could have different control levels such as ownership relationship between a user and a device, or the same level of privileges such as peer-to-peer relationship between two sensors. The management of hierarchical relationship is vital for security applications in IoT, only upper-level entities in the control hierarchy should have the access to the designated low-level entities.

Functional relationship \cite{khaled2018framework}: defines the relationship between the functions of the two IoT entities. For instance, the co-work relationship between two sensors that are performing the same task is a functional relationship. The functional relationships are used to manage the tasks assigned to the network entities, computational offload in Fog and Edge computing uses functional relationships between the devices to perform computational and storage tasks.

Spatial relationship \cite{SahraouiDhelim2016}: defines the spatial relationship between IoT entities, for instance in a smart home environment, two sensors located in the same room are connected by the co-located relationship \cite{dhelim2018cyber}. The spatial relationship could be expressed relative to a given reference, or absolute spatial relationship and expressed by a positioning system such as GPS.

Temporal relationship \cite{SahraouiDhelim2016}: defines the temporal relationship between two IoT entities or events related to these entities. For example, the fire alarm's start warning event is triggered after the fire sensor detects a fire. After, before, when and any relationship that defines time events are defined as temporal relationships.

Social relationship \cite{Dhelim2020entity}: defines a social relationship between two IoT entities or related entities in the control hierarchy. For example, the friendship relationship of two users. The social relationships between devices are the main motivation of the Social IoT (SIoT) paradigm \cite{Atzori2014}.

\begin{table}[htb!]
	\centering
	\caption{Relationship examples in each IoT layer}
	\label{relations_table}
	\begin{tabular}{|l|c|c|c|c|} \hline 
		\textbf{IoT layer} & 	\textbf{\makecell{Relation source}} & 	\textbf{Relation Type} & 	\textbf{Space} & 	\textbf{\makecell{Cyber \\mapping}} \\ \hline 
		\multirow{1}{*}{\parbox{0.1\linewidth}{\vspace{0.3cm} \makecell{Sensing \\ layer}}}
		& 	Co-sensing & 	Functional & 	Physical & 	CPS \\ \cline{2-5} 
		&  Node-Sink & 	Hierarchical & 	Physical & 	CPS\\ \cline{2-5}
		&  Sensor-Event & 	Temporal & 	Physical & 	CPS\\ \cline{2-5}
		&  Sensor-Location & 	Spatial &	Physical & 	CPS\\ \hline
		
		\multirow{1}{*}{\parbox{0.1\linewidth}{\vspace{0.3cm} \makecell{Network\\layer}}}
		& 	P2P & Hierarchical & 	Physical & 	SDN\\ \cline{2-5}
		&  Client-Server & 	Hierarchical & 	Physical & 	SDN\\ \cline{2-5} 
		&  M2MC & 	Functional & 	Physical & 	SDN\\ \cline{2-5} 
		&  Position-Routing & 	Spatial &	Physical & 	SDN\\ \hline 
		
		\multirow{1}{*}{\parbox{0.1\linewidth}{\vspace{0.2cm}\makecell{Application\\layer}}} & 	User-Account & 	Hierarchical & 	Social    & 	CPSS\\ \cline{2-5} 
		&  User-User & 	Social & 	Social & 	CPSS\\ \cline{2-5} 
		&  User-Data & 	Hierarchical & 	Social & 	CPSS\\ \hline 
	\end{tabular}
\end{table}

The relationship that holds among network entities in the physical space are mapping into the cyber space by a cyber-physical system (CPS) \cite{dartmann2019big}, similarly the application represented as entities in the social space are mapped to the cyber space using a cyber-physical social system (CPSS) \cite{zeng2020survey}. The relationships among various entities in the physical and social space are represented as cyber entities in the cyber space as shown in Figure \ref{cyber_mapping}. In Tabel \ref{relations_table}, we list examples of these relationships in each layer in the IoT architecture.

All these relationships of different types among connected objects bring to the creation of a social network which will contribute to the realization of the ASI. It is a social network where the objects are strictly interconnected to highlight key connections that are exploited for the realization of future communications and applications. This is a live network, as the observed nodes activities and profiles change over time also due to the varying applications used by the humans. Indeed, the creation of these links are triggered by the activities performed by the humans which are more and more frequently monitored by either personal devices or devices which are installed in the surroundings to improve the environment smartness. These devices are also involved in the execution of automatic tasks without the involvement of the humans, such as the setting of the working parameters of the appliances at home or the control of energy management systems in complex buildings. Accordingly, they cooperate with other objects and services on behalf of the humans and record the resulting interactions.

A major advantage of the resulting social network is that it fosters future interactions among all the peers (humans, devices and services) by providing a navigable network, i.e., it contains short paths among all (or most) pairs of nodes. However, to achieve this objective the resulting network is created to maintain certain properties, which are expressed mainly in terms of degree of local connectivity, size of the giant components, absence of isolated groups of peers. Accordingly, in the past major works have been proposed for the creation and update of a network with these characteristics, so as to select the best links among the possible options and to remove those that do not provide an important contribution of the navigability while increasing complexity \cite{Nitti15,Arch19}. Additionally, the devices social networks may contribute to the ASI by providing key data for the evaluation of the trust level among devices, which is extremely important when two or more peers interact for the first time. Indeed, friend devices can be inquired about the reliability of another node from which a service has to be requested. The inquired devices may provide feedback on past transactions and through these it is possible to infer the reliability of the target peer \cite{Jafarian2020}. Past works have also proposed the creation of objects networks which are devoted to some specific domains, as it is the case of the Social Internet of Vehicle (SIoV) \cite{Al15}. It relies on the VANETs technologies for the creation of a vehicular social network platform on the basis of vehicle-to-vehicle, vehicle-to-infrastructure and vehicle-to-internet communications. The SIoV system exploits the resulting network among physical components to foster the establishment of different types of communications and store key information (e.g., safety, efficiency, and infotainment messages) for different use cases for the intelligent transport systems (ITS). Still in the vehicular communications domain, social-driven clusters are created to implement physical and logical topologies when multi-hop wireless communications are needed among moving vehicles \cite{Mag16}. Observation of past encounters among nodes are also used in social-aware routing solutions, as in \cite{Mei15}, where a routing algorithm is proposed, which presents also the benefit of being stateless. Delay tolerant networks also may exploit the “small world” properties that characterize the social networks especially when addressing routing problems \cite{Wei15}. Social properties of mobile users are also addressed in \cite{Zhao17}, to the purpose of improving device-to-device multicast communications performance in terms of throughput while guaranteeing fairly channel allocation to different multicast clusters in radio networks.

\section{AI and ASI}
\label{sec.3}
The social aspect of human intelligence and the relationship between intelligence and social relationships have been discussed extensively since the ancient philosophy age.  Many researchers have discussed the importance of the social factor in human societies for centuries. Quoting from \cite{aronson2018social}, "Man is by nature a social animal; an individual who is unsocial naturally and not accidentally is either beneath our notice or more than human. Society is something in nature that precedes the individual. Anyone who either cannot lead the common life or is so self-sufficient and therefore does not partake of society, is either a beast or a god?" (Aristotle, Politics, c. 328 B.C.). The physical, social, and thinking dimensions of the living being and the relationship between these dimensions have been studied in ancient and modern philosophy \cite{Ning2016}. When Information and Communication Technology (ICT) emerged, the development of computing and communication technologies was always centered around the physical, social and thinking spaces of the users. In the physical space, with the fast development of robots in the last century, robotics researchers have always wanted to embed robots with human-like features, including physical appearance. They have succeeded to shape the robots in the form of the human body, and the result was humanoid robots \cite{Hirose2007}.  Moreover, with the recent advances in wearable sensors and mobile devices, the physical orientation of the computing paradigm has changed to become ubiquitously present with the users, and the result was the new user-centered trends in computing such as the Internet of People \cite{Conti2017}. In the thinking space development, computing systems as robots and devices were enabled by AI technologies to empower them with human-like intelligence and knowledge reasoning. That led to the emergence of various brain-related study fields that aim to mimic the human brain and thinking, such as AI. Besides, some researchers have even tried to physically merge machine intelligence with human intelligence to benefit from hybrid Human-Artificial Intelligence \cite{Chen2020}. In the social space development, devices and robots were assigned with social properties such as personality traits and artificial emotions and feelings \cite{CAI2020}.

\begin{figure}[htb!]
	\centering
	\includegraphics[width=0.8\columnwidth]{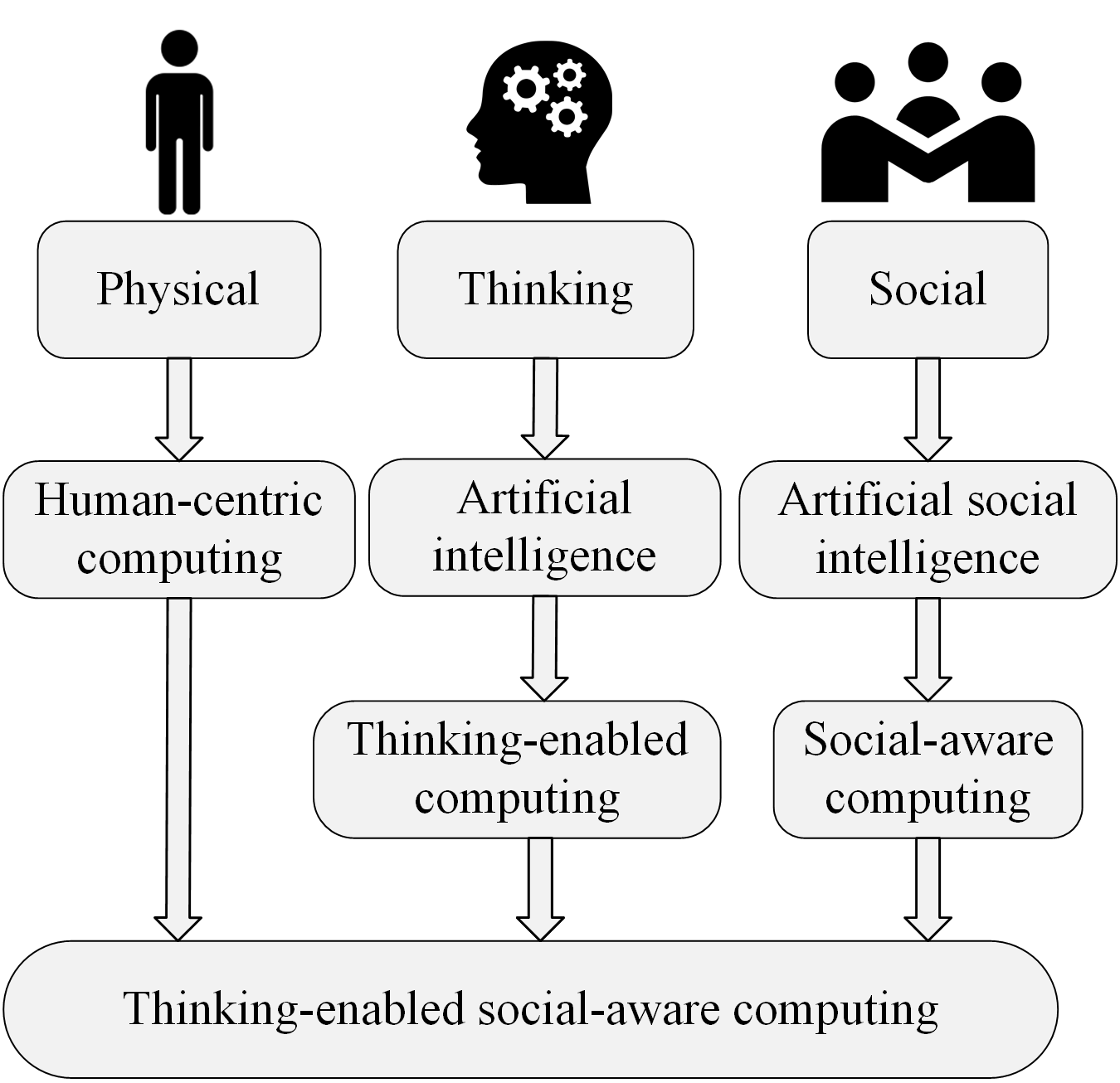}
	\caption{The development of physical-social-thinking spaces}
	\label{spaces}
\end{figure}

With the still ongoing AI revolution, AI is reshaping the future of different domains ranging from smart healthcare and smart services to industrial applications in supply chain and energy management \cite{jeschke2017industrial}. The future smart city roads will be crowded with humanoid robots, self-driving vehicles and smart delivery drones. Also, the future AI-enabled robots and machines will live side by side with us. Robots will share sidewalks with human pedestrians and socially interact with us. Therefore, in general, robots and smart devices should learn how to socially integrate with our society. For example, delivery robots must not interfere in people's personal space and should understand the social context during the delivery. While the development of AI is going at an unprecedented speed, the development of social-aware techniques is still at the early stages, and if this development trend continues in the current direction, we will end up with machines thinking-intelligent unsocial machines and devices. To avoid this scenario, machines' thinking abilities should be developed side-by-side with their social integration abilities, which required the tight coupling of AI techniques to be merged with the machines' social properties and social context. That resulted in the need for artificial social intelligence to be developed and tightly coupled with the conventional thinking AI. Figure \ref{spaces} illustrates the development direction of the future smart devices that would be able to leverage thinking intelligence and social intelligence as well.

\section{Social relationships exploding challenges}
\label{sec.4}
With the proliferation of social relationships, the IoT network faces new challenges at the unit IoT level and the ubiquitous IoT level. In this section, we present some of the challenges related to the social relationships explosion problem.
\subsection{Social Big Data}
The amount of generated data in each application unit of IoT is the scale of Terabytes (TB), not to mention the size of data generated by the ubiquitous IoT network \cite{ning2013unit}. According to the recent report of the International Data Corporation (IDC), by 2025 there will be 41 billion connected devices to the ubiquitous IoT network, and the amount generated data is estimated as 80 zettabytes (ZB) \cite{IDC2019}. The management of such a huge amount of data at the unit IoT level is an exceedingly challenging task, as the computational and communication resources in the unit IoT is very limited. The data management at the unit IoT level involves many pre-processing and filtering tasks, such as data aggregation and data compression.  At the ubiquitous IoT level, data management is even more challenging to aggregate and combine the generated data from heterogeneous IoT units. In addition to the classic big data challenges, the social relationships exploding will cause more complicated data management challenges. To understand and take advantage of the social relationships among users of social-aware unit IoT applications, the social knowledge of the local unit IoT is complemented with external data such as open linked data and knowledge graphs. The data linking of the social-aware applications with an external data source and the requirement of frequent updates will pose the conventional big data challenges in a more severe form \cite{Zhou2018}. AI techniques have been widely applied to tackle data management problems in all layers, from the physical layer to the application layer \cite{sun2016internet}. However, in the context of social-aware AI, the system should leverage the social dimension of the data and extract knowledge at all the data abstraction levels. That requires the system to change the way of data processing, from conventional data processing to more socially-aware data processing. As for socially-aware IoT applications, the challenge is not sensor and actuator management, nor the management of computing and communication, not even data management and manipulation in the physical sense. The real challenge is how to make sense of social data.
\subsection{Social feature processing}
The physical and emotional status of the users during a social relationship can unveil the hidden semantic and latent social feature of the social interaction that is taking place. With continuous development in the field of natural language processing and human-machine interaction, modern computing systems could recognize the textual as well as vocal communication of user-to-user or user-to-device interaction. Furthermore, the emerging of a new social computing paradigm had enabled machines to comprehend many social aspects of the user's social features. These include Affective Computing \cite{Poria2017}, the field that encompasses the development and design of devices and systems that can capture, process and detect human effects, and Personality Computing \cite{vinciarelli2014survey}, the study field that aim to integrate human personality traits in the computing systems, and many other social-related computing fields such as Sentiment Analysis \cite{Cambria2016} and Trust Computing \cite{Sherchan2013} to name a few.  However, these social computing technologies are not developed enough to comprehensively understand all the social features that semantically enrich social communication. With social relationships explosion, the challenge is to integrate these social feature processing technologies near end-user devices, that is because some of these technologies require low response time, for example integrating real-time facial affects recognition at a smart home camera. Social feature processing technologies will cause proliferation of social-enabled devices connected to the IoT network, which will cause additional communication overhead in these devices.
\subsection{Social context awareness}
The social internet of things (SIoT) aims to leverage the social dimension of smart objects to improve network navigability and device reachability, turning the network devices from a smart object into social-aware smart objects \cite{Atzori2014}. That will enable the smart objects to integrate into the social networks and improve its device-human interaction capabilities, such as the social-aware speaking objects \cite{Lippi2018} that communicate with humans using argumentation to show how various forms of human dialogue naturally fit coordination and cooperation requirements of the SIoT. To ensure the social context awareness of smart objects, the SIoT devices should be connected with social networking platforms, such as Facebook and Twitter, and with a personal social knowledge graph that maintains the knowledge about the social context of the users. With the massive social data generated from the social relationships explosions among users and devices, maintaining social context-awareness becomes an extremely challenging task, from the computational capability perspective as well as information semantic reasoning perspective. Furthermore, with the near future deployment of 5G networks, the social object will struggle to extract the social context from such a fast communication \cite{Al-Turjman2019}.
\subsection{Social data privacy}
Dealing with the social relationship explosion requires that the SIoT devices share the captured social features of the users with local as well as external applications to customize the offered services depending on the social context of the users. The social properties of users are sensitive information. In case of a network intrusion or data leaking, these social properties and social context data can be used to launch a socially engineered attack on the SIoT network \cite{Salahdine2019}. With the enormous number of SIoT connected devices, the management of multiple access levels and data sharing privileges become extremely challenging. The implementation of social-privacy preserving schemes that can protect the social properties of social relationship explosions is one of the main requirements of the future SIoT network.

\section{Computing and communication solutions to support IoT-enabled ASI}
\label{sec.5}
With the advances of social-aware computing and communication techniques, computing and communication tend to converge toward joint computing-communication social intelligence \cite{Bouras2020}.  In this section, we review some of the social-aware computing and communication techniques that address some of the challenges of SIoT and the convergence of social-aware computing and communication.
\subsection{ASI in computing} 
ASI-based machine learning and deep-learning techniques can be applied in many social computing tasks, ranging from social data preprocessing and feature extraction to service recommendations and application customization. In the feature select and classification task, ASI is applied to extract either raw features from social data, such as social feature selection \cite{IyapparajaM2020,jaysari2021}, or to classify, generate or detect contextual meaningful social cues, such as text sentiment classification \cite{Jiang2020}, automatic personality recognition \cite{ning2019personet},  natural language generation \cite{Jiang2020}, and user interest detection \cite{Dhelim2020,Dhelim2020compath}. With the applications and services computing tasks, ASI computing techniques are used to filter the offered services and match the right user with the right service according to the user's social properties and social context, such as social-aware service recommendation \cite{Khelloufi2020}, social-aware product recommendation \cite{metainterest} and social trust management \cite{SenthilKumar2020}. Table \ref{computing_table} lists some of the common computing tasks in SIoT and the literature of social-aware solutions.
\begin{table}
\centering
\caption{Common computing tasks in SIoT}
\label{computing_table}
\begin{tabular}{|m{0.25\columnwidth}|m{0.15\columnwidth}|m{0.35\columnwidth}|} \hline 
	\textbf{Social computing tasks} & \textbf{Publications} & \textbf{Task description} \\ \hline 
	Trust management in SIoT & \cite{SenthilKumar2020,Azad2020,Jayasinghe2019,Wei2020,Truong2017,Son2020,Jafarian2020,Nitti2014} &Establishing trust between SIoT devices for social clustering and social community detection\\ \hline 
	Social features extractions and classification & \cite{jaysari2021,IyapparajaM2020,Ma2019,Lakshmanaprabu2018} & Assigning social properties such as personality traits, mood, emotions and interest to SIoT users\\ \hline 
	Social relationships management in SIoT & \cite{Girau2017,Rajendran2020,Jung2018,Loscri2019,Ali2018,Atzori2019,aljubairy2020} & Maintaining the user-user, user-device and device-device social relationship history and logging the relationship properties and preferences \\ \hline 
	Social-aware recommendation system in SIoT & \cite{Khelloufi2020,Wei2020a,Cheng2019,Saleem2016,Chen2016,Lye2020} & Services recommendation and content customization based on the social properties of SIoT entities\\ \hline 
	Security and privacy in SIoT & \cite{Wu2015,Xia2020,Xia2020a,Zhang2020,Shen2018} & Securing and preserving the privacy of social data and social context\\ \hline 
\end{tabular}
\end{table}

\subsection{ASI in IoT communication}
ASI plays an important role in IoT communication, as objects can have social relationships established using matched or different communication technologies and across different IoT platforms \cite{Farris2015}. Managing and storing these relationships is usually done in the cloud. With the explosion in social relationships, the cloud's centralized server faces some significant challenges. The objects are far from the data centers hosted in the cloud, which led to delays, insufficient communication infrastructure and management bottleneck. IoT end devices are resource-constrained in general, which makes handling the sociality procedures add more overhead to the computing, communication and storage resources. Therefore, some techniques and paradigms are applied to mitigate the effects caused by the social relationships exploding and centralized cloud-based IoT.

\subsubsection{Edge Computing}
While the computation power and storage capacities resources are usually ubiquitous, sensing data, such as healthcare data, smart home data, or even the personal activities data, are correlated to physical locations. Using centralized cloud paradigms makes the data being transmitted to the cloud servers before returning to users. Therefore, the distributed cloud was introduced to make the processing of sensing information closer to the end devices, which results in reducing network latency and traffic congestion \cite{naouri2020}. Even though the distributed cloud has improved the whole system performance, there was a trend to push the services closer and closer to the users, and thus the edge computing concept comes into existence. When users request information, the requests will be processed locally \cite{Chiang2016}. So, the edge nodes need to be carefully designed to satisfy the services' requirements \cite{Ning2020}. Some projects already integrated the SIoT concepts with edge computing by creating virtual images to run social functionalities of the physical smart devices and then transfer them to the edge nodes. These images are like profiles that include information about the physical devices' capabilities, resources, and permissions \cite{Farris2015}.

\subsubsection{Mobile Edge Computing (MEC)}
It is also a paradigm that deploys resources at the mobile network edge on the base station. It handles mobile cloud computing requests and offers context and location awareness services. The main concept is improving the network performance, reducing unnecessary network traffic and increasing the throughput while replying to the users' requests \cite{Beck2014}. AI on chips has produced significant expansion in SIoT over recent years. It allows SIoT devices with social relationships to create groups that can collaborate to identify and handle some tasks locally without interference from remote servers. MEC is a common way to process social data locally, which reduces communications and computation overhead among SIoT devices and networks. The created social groups can process users' requests within the MEC networks that also have other techniques such as data aggregation to reduce the transmitted data amount effectively \cite{Ning2020}.

\subsubsection{Network Abstraction}
The cloud-based IoT cannot easily satisfy the requirements of scalability, big data, and mobility simultaneously. There are some limitations in the network architecture and protocols that make them, in some cases, not qualified to run real-time and delay-sensitive applications. To help address these issues, network abstraction, including some techniques, such as SDN and NFV, was presented in the network structure.

Software-Defined Networking (SDN) defines network behavior using the software. It mainly separates control and forwarding planes by building a model of forwarding plane and add some protocols for applying the control and network configuration \cite{Haleplidis2015}.

Network Function Virtualization (NFV) uses software virtualization instead of networking devices. It separates software from hardware, making it possible for the network functions to run on general-purpose hardware instead of using dedicated hardware devices, such as network switches, network routers, firewalls, and other devices \cite{Haleplidis2015}.

SDN and NFV are used to provide cloud services with low latency and high throughput \cite{Farris2015}. Moreover, putting virtual objects near their physical entities helps improve the use of resources. Routers in Internet service providers usually have good computation and storage resources for creating virtual machines to run user applications. Some projects, such as the INPUT Project \footnote{www.inputproject.eu}, exploit SDN and NFV \cite{Jain2013} to apply services dynamically in the network edge and make the computing and storage resources distributed through the network in addition to moving the services to users nearby \cite{Farris2015}.
\subsubsection{Device to Device (D2D) communication}
D2D communication has been developed to meet the increased interest in transferring the data locally at high data rates. It enables direct connections with nearby users without the help of a higher-level device \cite{Nitti2019}. SIoT involves things and people in networks and establishes connections guided by social relationships and controlled by the device owners' rules \cite{Nitti2019}.

The relationship between D2D communications and SIoT has recently attracted the attention of researchers and developers, where social networking can increase D2D communication effectiveness and performance. This combination can also achieve high throughput, better data rate, low latency and lower power consumption. Social networking can help D2D communications find and benefit from relationships of social network users and improve its competence when it is built using proximity information. To do that, SIoT helps relay discovery and peer selection, primarily depending on the neighbor discovery process, which can be time and energy-consuming without social information in the network \cite{Ahmed2017}.

\subsubsection{Computing and communication convergence and ASI}
The computing processes, control, and storage were recently pushed from centralized cloud computing to the network edge to allow real-time, critical and computation-intensive applications to run on the resource-constrained devices. Balancing between communications and computing results in various new designs ranging from computation offload techniques to network architectures \cite{Mao2017}.

SIoT can make collaboration among its users to effectively reduce communications. On the other hand, more AI computations on the central SIoT are required to understand the environmental events and consider that neighbors' devices are not always connected to the same social network. Also,  friends in social networks can be distant from each other, which requires multi-hop transmissions.

\section{ASI use case scenario} 
\label{sec.r2}
The necessity of applying ASI in IoT is better illustrated through a concrete use case scenario. let's consider an ASI-enabled smart home scenario, where the IoT devices within the smart home and the residences' smartphone applications are connected to ASI-enabled smart home processing unit. Adam is a resident of an ASI-enabled smart home, before sleeping he posts on Twitter "so excited for tomorrow's job interview". As the smart home is ASI-enable, by processing this tweet, it deduce Adam's social context and add this event to tomorrow's schedule. Adam had forgotten to set the alarm for tomorrow's interview. Fortunately, the ASI-enabled smart home automatically set the alarm, print Adam's required documents and prepare his self-driving car for the trip. Upon finishing his interview, Adam feels hungry and asks his ASI-enabled voice assistant device to recommend a nearby restaurant. Without considering Adam's social context, a conventional voice assistant device would recommend the nearest restaurant. But since Adam's device is ASI-enabled, it filters nearby restaurants that best suits his social context, and by analyzing his social network data find out that his high school friend Bob, which he did not meet for 10 years, happen to be having lunch in a nearby restaurant, that is because Bob have shared his location a few minutes ago. After having lunch with his high school friend, Adam decided to head back home. Adam was surprised that his ASI-enabled self-driving car took a relatively longer path, however Adam later on realized that the obvious shortest path has been congested for hours due to some organized even within that path, his ASI-enabled car predicted that by analyzing the social media content at the population-level.

\section{ASI in IoT applications} 
\label{sec.6}
The adaptation of ASI will revolutionize the IoT applications and services from different aspects. In this section, we present some of the ASI-enabled IoT applications.

\subsection{Mental healthcare in IoT}
The ambient connectivity provided by the IoT network can provide instant and ubiquitous access to healthcare facilities and services. Healthcare-related IoT applications have many advantages over traditional healthcare services. As they allow for remote patient management, all the healthcare protocols, from the diagnosis to the treatment, can be coordinated by telemedicine through IoT network \cite{Qadri2020}.  However, the usage of IoT applications in the healthcare sector is still limited in treating conventional diseases, to things like heartbeat and blood pressure monitoring \cite{Moghadas2020}. Additionally, mental and behavioral disorders applications require more than reactive sensing and monitoring of the patient's physical status. They need to predict, prevent and proactively engage with the mental conditions of the patient. Here comes the role of ASI, as the reasoning of ASI-enabled healthcare application is empowered by the social context of the patient, it can offer psychological support to the user without even explicitly being commanded. For instance, depression and loneliness among the elderly have reached an unprecedented level. They are considered two of the most common mental health disorders for the elderly, especially in developed countries \cite{Dury2014}.

With the help of ASI, the elderly are surrounded by social-aware machines embedded with human-like personality traits designed especially to be harmonic with the aged user's personality traits \cite{Kachouie2014}. One of the advantages of ASI healthcare application compared to traditional human-assisted healthcare is the social context awareness of ASI, which enables machines to understand the social context of every patient. ASI-based children's healthcare is another promising application. ASI-enabled humanoid robots are developed to aid children to subsist the painful medical procedures. Many previous studies have shown the importance of social-enabled psychological systems in reducing the pain and stress of the patients \cite{foster2020}. These promising applications of ASI must be accompanied by the risk awareness of the treated medical conditions. In other words, the ASI agent must be cognizant of and guarantee adequate risk management \cite{Hague2019}.

\subsection{Intelligent transportation and ASI}
The application of AI has revolutionized intelligent transportation systems in the last decade. It has been applied to solve a wide range of traffic-related problems, such as traffic communication systems \cite{Lv2020} and traffic estimation \cite{Khan2017a}. The integration of the commuters' social context with intelligent transportation systems using ASI will yield even more efficient traffic-related applications.

Traffic path planning and routing are two of the most challenging traffic problems because bad traffic routing is the main cause of traffic congestion \cite{Aung2020}. However, the legacy AI path planning systems consider all the vehicles equally, focus only on the desired destination and schedule the routes accordingly. Some other solutions have tried to optimize the path planning computations using a distributed server architecture \cite{Zhang2018}. While ASI includes the social context of the drivers, such as social media's events, their previous traveling history and the logged routes preferences \cite{Arooj2020}.  Driver-less vehicles are one of the most anticipated technologies in the future of intelligent transportation. While conventional driver-less vehicles are completely dependent on AI to navigate the roads and perform their assigned task, the need to be enhanced using ASI is not just a complimentary improvement. ASI can empower the vehicles with social rules and common sense conventions that enable them to cooperate with other vehicles' drivers. Integrating the driver's social context also helps secure communication among vehicles by maintaining a social trust and reputation system \cite{aung2018accident}.

\subsection{ASI and smart city}
Millions of people will inhabit the future large smart cities. The applications of unit IoT level, such as smart home or smart community, are expected to generate social data that can be semantically meaningless at the unit IoT scale but meaningful at the ubiquitous IoT level. The ASI is applied to the generated social data from the unit IoT level to achieve collective intelligence using crowed sensing and crowed computing techniques. The conventional AI techniques focus on solving the problems that occur while dealing with large scale smart city challenges, such as large scale video management \cite{song2017smart} or massive devices multiple access \cite{Duan2019}. 

ASI is leveraged to solve social-oriented large-scale challenges. For instance, the social media content of the smart city residents is a rich source of social information that can help optimize smart city services. ASI-enabled social media analysis and the analysis of crowd sensed location data and other data about the social context of the smart city inhabitants can be used to detect and early prevent a terrorist plot that targets the smart city. Another prominent application of ASI in the context of smart cities is the systematic analysis of social media triggered events that can have physical consequences on real-life, such as cyberbullying incidents that might be followed by physical attacks or suicide incidents.

\section{Conclusion}
\label{sec.7}
The unprecedented proliferation of social relationships among IoT entities has led to computing and communication bottlenecks known as social relationships explosion. In this paper, we have discussed the problem of social relationships explosion in IoT and show that the emerging artificial social intelligence has the potential to tackle the social relationships explosion problem. Unlike conventional artificial intelligence, artificial social intelligence is integrated with the computing and communication techniques, which enable it to deal with the social relationships explosion from a social computing perspective. The social-centered convergence computing and communication will enable the IoT devices to take advantage of social context to improve the offered services and customize the delivered content.

	\section*{Acknowledgment}
This work was supported by the National Natural Science Foundation of China under Grant 61872038.

	\ifCLASSOPTIONcaptionsoff
	\newpage
	\fi

	
	
	\bibliographystyle{IEEEtran}
	
	\bibliography{refs}
	%

	%
	
\vskip 0pt plus -1fil	

\begin{IEEEbiography}[{\includegraphics[width=1in,height=1.25in,clip,keepaspectratio]{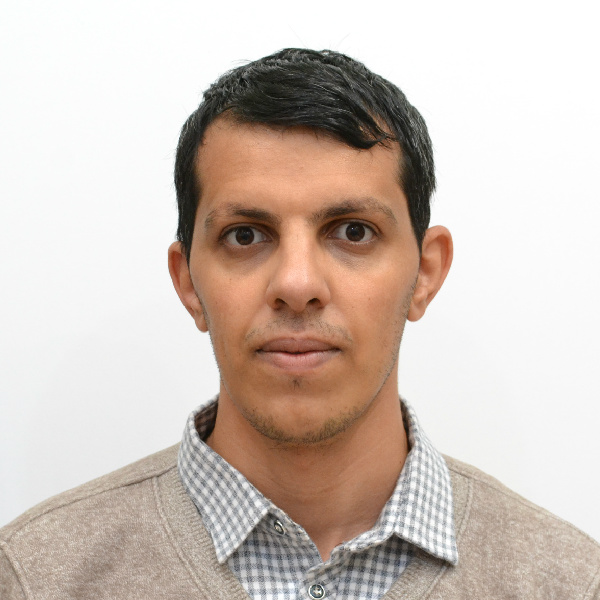}}]{Sahraoui Dhelim}
	Received his B.S. in Computer Science from the University of Djelfa, Algeria, in 2012 and his Master degree in Networking and Distributed Systems from the University of Laghouat, Algeria, in 2014, and PhD in Computer Science and Technology from University of Science and Technology Beijing, China, in 2020. His current research interests include Social Computing, Personality Computing, User Modeling, Interest Mining, Recommendation Systems and Intelligent Transportation Systems.
\end{IEEEbiography}

\vskip 0pt plus -1fil

\begin{IEEEbiography}[{\includegraphics[width=1in,height=1.25in,clip,keepaspectratio]{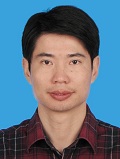}}]{Huansheng Ning}
	Received his B.S. degree from Anhui University in 1996 and his Ph.D. degree from Beihang University in 2001. Now, he is a professor and vice dean of the School of Computer and Communication Engineering, University of Science and Technology Beijing, China. His current research focuses on the Internet of Things and general cyberspace.
	He is the founder and chair of the Cyberspace and Cybermatics International Science and Technology Cooperation Base.
	He has presided many research projects including Natural Science Foundation of China, National High Technology Research and Development Program of China (863 Project). He has published more than 100+ journal/conference papers, and authored 5 books. He serves as an associate editor of IEEE Systems Journal (2013-2020), IEEE Internet of Things Journal (2014-2018),  steering committee member of IEEE Internet of Things Journal
	(2016-2020), and area editor (2020-now).
	
\end{IEEEbiography}

\vskip 0pt plus -1fil

\begin{IEEEbiography}[{\includegraphics[width=1in,height=1.25in,clip,keepaspectratio]{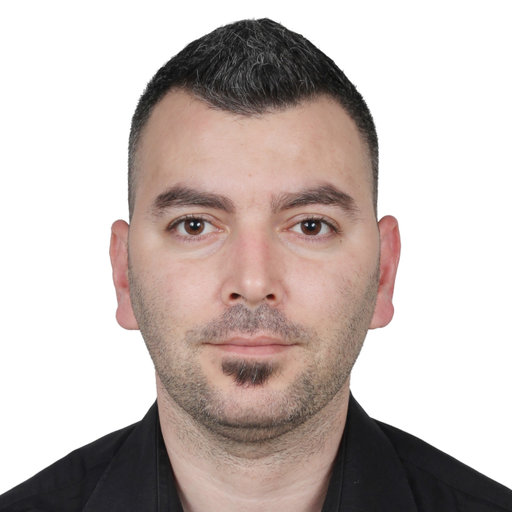}}]{Fadi Farha}
received his BS from the faculty of Informatics Engineering, Aleppo University, Syria. He did his MS degree and currently working toward a Ph.D.degree in the School of Computer and Communication Engineering, University of Science and Technology Beijing, China. His current research interests include Physical Unclonable Function (PUF), Security Solutions, ZigBee, Computer Architecture, and Hardware Security.
\end{IEEEbiography}

\vskip 0pt plus -1fil

\begin{IEEEbiography}[{\includegraphics[width=1in,height=1.25in,clip,keepaspectratio]{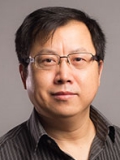}}]{Liming Chen}
	is a professor in the School of Computer  Science  and  Informatics  at  University  of  Ulster,  Newtownabbey,  United  Kingdom.  He  received his  B.Eng  and  M.Eng  from  Beijing  Institute  of Technology  (BIT),  Beijing,  China,  and  his  Ph.D  in Artificial Intelligence from De Montfort University,UK.  His  research  interests  include  data  analysis,ubiquitous computing, and human-computer interaction. Liming is a Fellow of IET, a Senior Member of IEEE, a Member of the IEEE Computational Intelligence Society (IEEE CIS), a Member of the IEEE CIS Smart World Technical Committee (SWTC), and the Founding Chair of the IEEE CIS SWTC Task Force on User-centred Smart Systems (TF-UCSS). He has served as an expert assessor, panel member and evaluator for UK EPSRC (Engineering and Physical Sciences Research Council, member of the Peer Review College), ESRC (Economic and Social Science Research Council), European Commission Horizon 2020 Research Program, Danish Agency for Science and Higher Education, Denmark, Canada Foundation for Innovation (CFI), Canada, Chilean National Science and Technology Commission (CONICYT), Chile, and NWO (The Netherlands Organisation for Scientific Research), Netherlands.
	
\end{IEEEbiography}

\vskip 0pt plus -1fil

\begin{IEEEbiography}[{\includegraphics[width=1in,height=1.25in,clip,keepaspectratio]{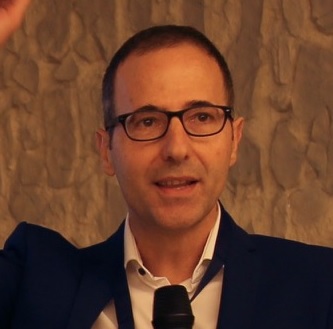}}]{Luigi Atzori} is Full Professor at the Department of Electrical and Electronic Engineering, University of Cagliari (Italy) and Research Associate at the Multimedia Communications Laboratory of CNIT (Consorzio Nazionale Inter-universitario per le Telecomunicazioni). Prof. Atzori received his Ph.D. degree in electronic engineering and computer science from the University of Cagliari in 2000. He spent the seven months from November 2003 to May 2004 at the Department of Electrical and Computer Engineering, University of Arizona, as Fulbright Visiting Scholar. L. Atzori research interests are in multimedia communications and computer networking and services in the Internet of Things and Social Internet of Things. L. Atzori is senior member of IEEE (since 2009) and has been the Steering Committee Chair of the IEEE Multimedia Communications Committee (MMTC) for the years 2014-2016. He has been the associate and guest editor for several journals, included: ACM/Springer Wireless Networks Journal, IEEE IoT journal, IEEE Comm. Magazine, the Springer Monet Journal, Elsevier Ad Hoc Networks, and the Elsevier Signal Processing: Image Communications Journal. Currently he serves in the editorial board of the following journals: Elsevier Digital Communications and Networks and IEEE Open Journal of the Communications Society. He served as a technical program chair for various international conferences and workshops, including ICC and Globecom workshops, ACM MobiMedia and VLBV. He served as a reviewer and panelist for many funding agencies, including FP7, Horizon2020, Cost Actions, Italian MIUR and Regional funding agency.
\end{IEEEbiography}

\vskip 0pt plus -1fil

\begin{IEEEbiography}[{\includegraphics[width=1in,height=1.25in,clip,keepaspectratio]{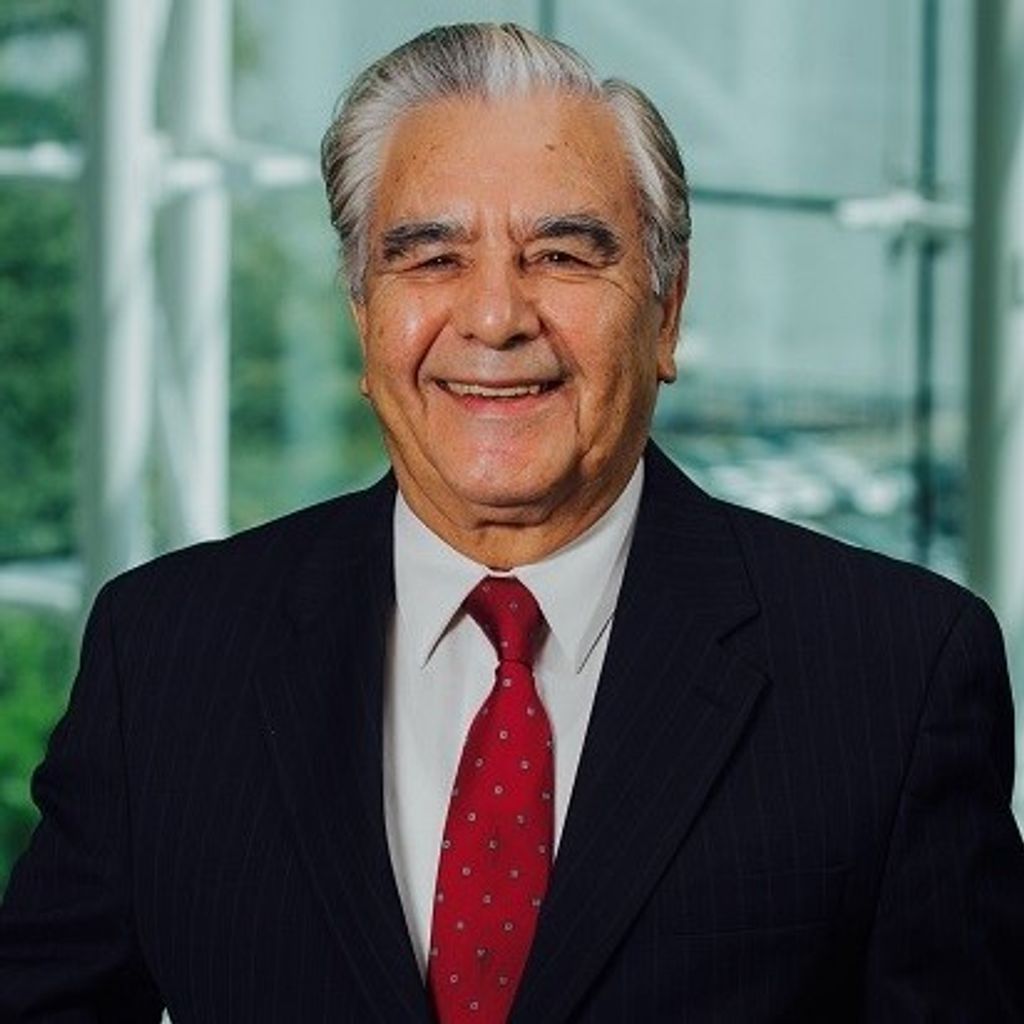}}]{Mahmoud Daneshmand} received his Ph.D and M.S. degrees in Statistics from the University of California, Berkeley; M.S. and B.S. degrees in Mathematics from the University of Tehran. He is currently an Industry Professor with the Department of Business Intelligence \& Analytics as well as Department of Computer Science at Stevens Institute of Technology, USA. He has more than 35 years of Industry \&University experience as: Professor, Researcher, Assistant Chief Scientist, Executive Director, Distinguished Member of Technical Staff, Technology Leader, Chairman of Department, and Dean of School at: Bell Laboratories; AT\&T Shannon Labs–Research; University of California, Berkeley; University of Texas, Austin; Sharif University of Technology; University of Tehran; New York University; and Stevens Institute of Technology. He has published more than 150 journal and conference papers; authored/co-authored three books. He is well recognized within
the academia and industry and holds key leadership roles in IEEE Journal Publications, Conferences, Industry IEEE Partnership, and IEEE Future Direction Initiatives. He is CoFounder and Chair of Steering Committee of IEEE IoT Journal; Member of Steering Committee of IEEE Transaction on Big Data; guest editor of several IEEE publications; CoFounder of the IEEE Big Data Initiative; and has served as General Chair, Keynote Chair, Panel Chair, and Technical Program Chair of many IEEE major conferences. He has given several Keynote speeches in IEEE as well as international conferences. He is an expert on Big Data Analytics with extensive industry experience including with the Bell Laboratories as well as the Info Lab of the AT\&T Shannon Labs – Research.
\end{IEEEbiography}

	
	

\end{document}